\documentclass[pra,twocolumn,showpacs,preprintnumbers,amsmath,linenumbers,amssymb]{revtex4}
\usepackage{amsfonts}
\usepackage{mathrsfs}
\usepackage{amsmath}
\usepackage{amssymb}
\usepackage{txfonts}

\usepackage{graphicx}
\usepackage{dcolumn}
\usepackage{bm}
\usepackage[dvipdfm]{xcolor}
\usepackage{subfigure}

\begin{document}



\title{Active Ion Optical Clock }

\author{Wei Zhuang, Tonggang Zhang, Jingbiao Chen$^\dag$}

\address{Institute of Quantum Electronics, and State Key Laboratory of
Advanced Optical Communication System $\&$ Network School of
Electronics Engineering and Computer Science, Peking University,
Beijing 100871, China. }


\begin{abstract}
In this paper, we propose a scheme of active ion optical clock
with detailed pumping method, lasing states, output power, linewidth and light
shift. Using $^{171}$Yb$^{+}$ ions in a Paul Trap we propose to utilize a
Fabry-Perot resonator to realize lasing of active optical frequency standards.
The quantum-limited linewidth of active $^{171}$Yb$^{+}$ ion optical clock
is narrower than 1 mHz. Based on the mechanism and advantages of active optical clock at ion optical clock
transition frequency, this new laser light source as a stable local oscillator, will be beneficial to
the single-ion optical clock, which currently is the most accurate clock.
\end{abstract}

\pacs{06.30.Ft, 42.55.-f, 37.30.+i}

\maketitle

\section{INTRODUCTION}

Optical clocks, with natural linewidth at the millihertz level, have demonstrated great
improvements in stability and accuracy over the microwave frequency standards.
The research on optical atomic clocks have achieved remarkable progress in the past
several years especially in single-ion optical clocks.
All-optical atomic clock referenced to the 1.064 petahertz transition of a single trapped $^{199}$Hg$^{+}$
ion has been realized in 2001~\cite{Diddams}. Two Al$^{+}$ ion optical clocks, operated at
$^{1}$S$_{0}$ to $^{3}$P$_{0}$ clock transition with frequency near 1.121 petahertz and narrow natural linewidth of 8 mHz,
have been constructed with fractional frequency inaccuracy of the order of magnitude $10^{-18}$~\cite{Chou,Rosenband}.
Although these single-ion optical clocks have reached unprecedented stability and accuracy, there are still some
problems to be solved. The observed linewidth of the clock transition is limited by the linewidth of the
probe laser~\cite{Chou2,Jiang,Katori,Swallows}. Thus, narrow linewidth laser light source becomes the
key factor of the performance of single-ion optical clocks.

Since the proposal of active optical clock~\cite{Chen1,Chen2},
a number of neutral atoms with two-level, three-level and four-level at thermal,
laser cooling and trapping configurations have been investigated
recently~\cite{Chen1, Chen2, Zhuang1, Zhuang2, Zhuang3, Yu, Chen3, Meiser1, Meiser2, Xie, Zhuang4, Zhuang5}.
The potential quantum-limited linewidth of active optical clock is narrower
than mHz, and it is possible to reach this unprecedented linewidth since the thermal
noise of cavity mode can be reduced dramatically with the mechanism of
active optical clock~\cite{Chen1, Chen2, Zhuang1, Zhuang2, Zhuang3, Yu, Chen3, Meiser1, Meiser2, Xie, Zhuang4, Zhuang5}.
Therefore, a laser light source based on active ion optical clock
will be favorable to single-ion optical clocks.

One candidate is $^{171}$Yb$^{+}$ ion. With about $10^6$ $^{171}$Yb$^{+}$ ions in a Paul Trap~\cite{Casdorff},
the population inversion can be realized as shown in Fig.~1. The cooling laser and repumping laser at
$^{2}$S$_{1/2}(F=1)$ to $^{2}$P$_{1/2}(F=0)$ and $^{2}$S$_{1/2}(F=0)$ to $^{2}$P$_{1/2}(F=1)$
transitions are used as pumping laser. First, during the cycling $F=1$ to $F=0$ transition, there
is a branching probability of $6.6\times10^{-3}$ for ions decaying to $^{2}$D$_{3/2}(F=1)$
sublevel~\cite{Tamm1}. Second, when the ions that leaked to $^{2}$S$_{1/2}(F=0)$ are repumped to
$^{2}$P$_{1/2}(F=1)$, ions will decay to $^{2}$D$_{3/2}(F=1, 2)$ sublevels. Then the population
inversion is established between $^{2}$D$_{3/2}(F=1)$ and $^{2}$S$_{1/2}(F=0)$ sublevels. The ions at
$^{2}$D$_{3/2}(F=2)$ can be repumped with a 935 nm laser at transition from
$^{2}$D$_{3/2}(F=2)$ to $^{3}$D[3/2]$_{1/2}(F=1)$ and then decay to $^{2}$S$_{1/2}(F=1)$
by spontaneous transition very quickly. So this channel can be ignored when calculating the energy level
population. Small leakage channels to other
sublevels can be closed with additional repumping lasers if necessary.

\begin{figure}[!h]
\includegraphics[width= 6cm]{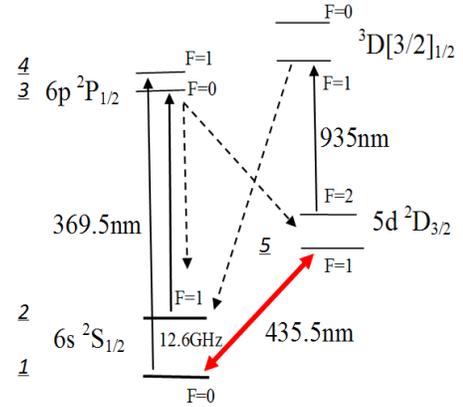}
\caption{(Color online) Scheme of $^{171}$Yb$^{+}$ ion laser
operating at 435.5 nm clock transition, pumped  with 369.5 nm and
935 nm transitions.} \label{fig:Fig1}
\end{figure}

The calculations of dynamical process of
population inversion of $^{171}$Yb$^{+}$ ions in a Paul Trap
and lasing process with a Fabry-Perot type resonator are presented in detail.
We calculated the quantum-limited linewidth of active $^{171}$Yb$^{+}$ ion optical clock and got it should be narrower than 1 mHz.
Other proper candidate ions for active optical clock and ion laser are also discussed and compared with
$^{171}$Yb$^{+}$ in this paper.

\section{POPULATION INVERSION OF IONS IN A PAUL TRAP AND LASING OF YB IONS ACTIVE OPTICAL FREQUENCY STANDARD}

Considering the spontaneous decay rate of $^{2}$P$_{1/2}$ is $2\pi\times23$ MHz, it is reasonable to set
the pumping rate~\cite{Tamm1} from $^{2}$S$_{1/2}(F=1)$ to $^{2}$P$_{1/2}(F=0)$ $\Omega_{23}=10^{7} s^{-1}$.
So the decay rate to $^{2}$D$_{3/2}(F=1)$ sublevel is $6.6\times10^{4}s^{-1}$. Then we can set the repumping rate of
$^{2}$S$_{1/2}(F=0)$ to $^{2}$P$_{1/2}(F=1)$ transition $\Omega_{14}$ to be of the order of magnitude $10^{5}s^{-1}$
and the repumping rate of $^{2}$D$_{3/2}(F=2)$ to $^{3}$D[3/2]$_{1/2}(F=1)$ sublevel to be $10^{3}s^{-1}$.
The effective power is reduced to one third if only
the 0-0 clock transition is effective. In order to depopulate ions at $^{2}$D$_{3/2}(F=1)$
($m_{F}=1,-1$) Zeeman sublevels, each transition from these two Zeeman sublevels
$^{2}$D$_{3/2}(F=1)$ ($m_{F}=1,-1$) to $^{2}$S$_{1/2}(F=0)$ is also coupled with one mode of ion laser
cavity.

Based on the assumption that the ions have been trapped in a Paul Trap, and the theories of the
interaction between the ions and light, the density matrix equations for $^{171}$Yb$^{+}$
ions interacted with the cooling light and the repumping light can be written in RWA-approximation as follows:
\begin{equation}
\begin{aligned}
\nonumber\frac{d\rho_{11}}{dt}&=-\Omega_{14}\rho^{I}_{14}+\Gamma_{41}\rho_{44}+\Gamma_{51}\rho_{55}-Kn\rho_{11}+Kn\rho_{55}\\
\nonumber\frac{d\rho_{22}}{dt}&=-\Omega_{23}\rho^{I}_{23}+\Gamma_{32}\rho_{33}+\Gamma_{42}\rho_{44}+\Gamma_{52}\rho_{55}\\
\nonumber\frac{d\rho_{33}}{dt}&=\Omega_{23}\rho^{I}_{23}-(\Gamma_{32}+\Gamma_{35})\rho_{33}\\
\nonumber\frac{d\rho_{44}}{dt}&=\Omega_{14}\rho^{I}_{14}-(\Gamma_{41}+\Gamma_{42}+\Gamma_{45})\rho_{44}\\
\nonumber\frac{d\rho_{55}}{dt}&=\Gamma_{45}\rho_{44}+\Gamma_{35}\rho_{33}-(\Gamma_{51}+\Gamma_{52})\rho_{55}+Kn\rho_{11}-Kn\rho_{55}\\
\nonumber\frac{d\rho^{I}_{14}}{dt}&=\frac{1}{2}\Omega_{14}(\rho_{11}-\rho_{44})+\rho^{R}_{14}\Delta_{1}-\frac{1}{2}(\Gamma_{41}+\Gamma_{42}+\Gamma_{45})\rho^{I}_{14}\\
\nonumber\frac{d\rho^{R}_{14}}{dt}&=-\rho^{I}_{14}\Delta_{1}-\frac{1}{2}(\Gamma_{41}+\Gamma_{42}+\Gamma_{45})\rho^{R}_{14}\\
\nonumber\frac{d\rho^{I}_{23}}{dt}&=\frac{1}{2}\Omega_{23}(\rho_{22}-\rho_{33})+\rho^{R}_{23}\Delta_{2}-\frac{1}{2}(\Gamma_{32}+\Gamma_{35})\rho^{I}_{23}\\
\nonumber\frac{d\rho^{R}_{23}}{dt}&=-\rho^{I}_{23}\Delta_{2}-\frac{1}{2}(\Gamma_{32}+\Gamma_{35})\rho^{R}_{23}\\
\nonumber\frac{dn}{dt}&=Kn(\rho_{55}-\rho_{11})-\Gamma_{c}n
\end{aligned}
\end{equation}

The subscript numbers of the density matrix correspond to different
energy levels as shown in Fig.~1. The diagonal elements mean the
number of ions in corresponding states and off-diagonal elements
mean coherence between two states. The Rabi frequency is
$\Omega_{23}=d_{23}\varepsilon_{1}/\hbar$, where $d_{23}$ is the
electric dipole matrix between the $^{2}$S$_{1/2}(F=1)$ state and
$^{2}$P$_{1/2}(F=0)$ state, $\varepsilon_{1}$ is the electric
strength of the cooling light.
$\Omega_{14}=d_{14}\varepsilon_{2}/\hbar$, with the electric dipole
matrix $d_{14}$ between the $^{2}$S$_{1/2}(F=0)$ state and
$^{2}$P$_{1/2}(F=1)$ state and the electric strength
$\varepsilon_{2}$ of the repumping light.
$\Delta_{1}(=\omega_{23}-\omega_{1})$ and
$\Delta_{2}(=\omega_{14}-\omega_{2})$ are frequency detunings of
cooling and repumping light on the transition frequencies. They are
both set to be 0. $\Gamma_{32}$, $\Gamma_{35}$, $\Gamma_{41}$,
$\Gamma_{42}$, $\Gamma_{45}$, $\Gamma_{51}$ and $\Gamma_{52}$ are
decaying rates related to the lifetimes of the corresponding energy
levels described in Fig.~1. $\Gamma_{c}$ is the cavity loss rate and
$n$ is the photon number if there exists a cavity. $K\sim
g^{2}t_{int}$ is the laser emission coefficient~\cite{An} where $g$
is ion-cavity coupling constant and $t_{int}$ is the interaction
time. Without the cavity, $\Gamma_{c}$ and $K$ are both 0. The
effective decay rate of the channel, from 6p $^{2}$P$_{1/2}$ to 5d
$^{2}$D$_{3/2}(F=0)$ then repumped with 935 nm laser to
$^{3}$D[3/2]$_{1/2}(F=1)$ and decay to $^{2}$S$_{1/2}$ by
spontaneous transition, is ignored in calculations since it is much
smaller than $\Gamma_{32}$.

For the convenience of reading, here we summarize all of the relevant parameters mentioned above
in Table~1 as follows.

\begin{center}
\textbf{Table 1. Parameters related to\\active ion optical clock}
\end{center}

\begin{center}
\renewcommand{\arraystretch}{1.5}
\begin{tabular}{|c|c|c|c|}
\hline
parameters & value & parameters & value\\
\hline
$\Gamma_{32}$ & $2\pi\times23$ MHz & $\Omega_{23}$ & $1\times10^{7}s^{-1}$\\
\hline
$\Gamma_{35}$ & $2\pi\times152$ kHz & $\Omega_{14}^{*}$ & $3.35\times10^{5}s^{-1}$\\
\hline
$\Gamma_{41}$ & $2\pi\times23$ MHz & $\Gamma_{c}^{**}$ & 500 kHz\\
\hline
$\Gamma_{42}$ & $2\pi\times23$ MHz & $K^{**}$ & 10$s^{-1}$\\
\hline
$\Gamma_{45}$ & $2\pi\times152$ kHz & $\Delta_{1}$ & 0\\
\hline
$\Gamma_{51}$ & $19$ Hz & $\Delta_{2}$ & 0\\
\hline
$\Gamma_{52}$ & $19$ Hz &    &  \\
\hline
\end{tabular}
\end{center}
\ \ \ \ \ \ $*$ Another value for discussion is $4.75\times10^{5}s^{-1}$.

\ \ $**$ Without the cavity, the value is 0.

\begin{figure}[!h]
\includegraphics[width=7.5cm]{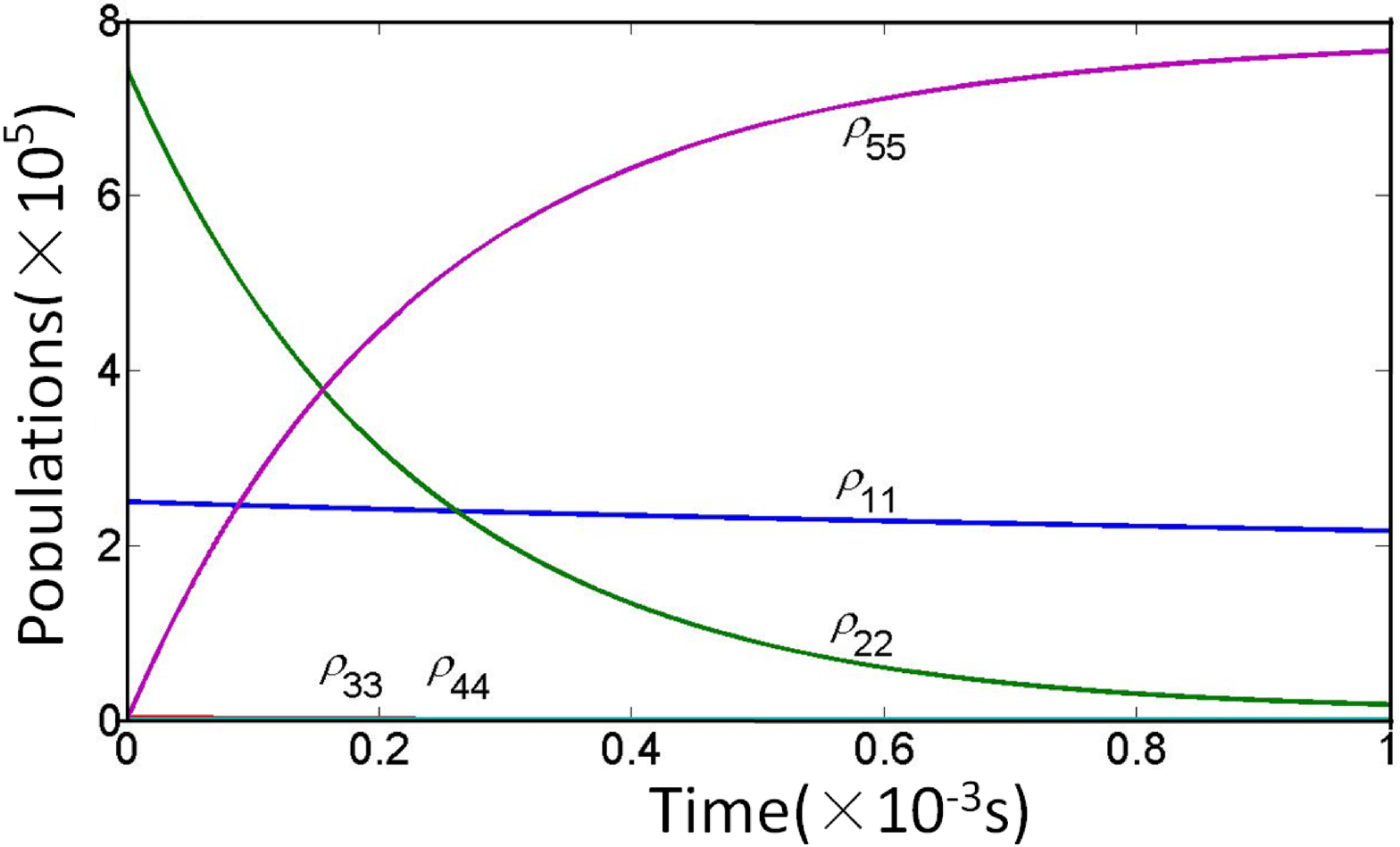}
\caption{(Color online) The dynamical populations with the Rabi
frequency $\Omega_{23}=10^{7} s^{-1}$ and
$\Omega_{14}=3.35\times10^{5} s^{-1}$.} \label{fig:Fig2}
\end{figure}

Without the cavity, the numerical solution results for the above equations are shown in Fig.~2. Assuming there are $10^{6}$ ions
in the Paul Trap and a quarter of them are in the $^{2}$S$_{1/2}(F=0)$ state at the beginning, the population inversion between
$^{2}$D$_{3/2}(F=1)$ and $^{2}$S$_{1/2}(F=0)$ states is built up at the time scale of $10^{-3}$s. From Fig.~2,
it is obvious that under the action of cooling laser, the number of ions in the $^{2}$S$_{1/2}(F=1)$ state decrease rapidly. The lifetime
of $^{2}$P$_{1/2}(F=0)$ and $^{2}$P$_{1/2}(F=1)$ states are much shorter than that of $^{2}$D$_{3/2}(F=1)$ state,
so ions accumulate in $^{2}$D$_{3/2}(F=1)$ state and the population inversion is built up.
If we keep other conditions unchanged, the number of ions at the steady-state in $^{2}$D$_{3/2}(F=1)$ sublevel is 10 times as
much as that in $^{2}$S$_{1/2}(F=0)$ sublevel for $\Omega_{14}=3.35\times10^{5}s^{-1}$, and 20 times for
$\Omega_{14}=4.75\times10^{5}s^{-1}$.

An optical resonant bad cavity, whose linewidth of cavity mode is much wider than the linewidth of gain,
could be applied to the lasing transition between the inverted states
$^{2}$D$_{3/2}(F=1)$ and $^{2}$S$_{1/2}(F=0)$, as the population inversion occurs. The oscillating process
could start up once the optical gain exceeds the loss rate. The corresponding density matrix
equations together with the equations of emitted photons($n$) from the total ions($N$)
inside the cavity can be written as above. We use a Fabry-Perot resonator with mode volume about
$3.3\times10^{-7} m^{3}$ and the cavity loss rate $\Gamma_{c}=500$ kHz. The laser emission coefficient is $K=10 s^{-1}$.

Fig.~3 describes the solution of the photon number equation. From Fig.~3 we can conclude that
a stable laser field has been built up within laser cavity on condition that the population inversion is preserved.
The steady-state value of photon number $n$ is 160 and
the power of this $^{171}$Yb$^{+}$ ion laser lasing from $^{2}$D$_{3/2}(F=1)$ to
$^{2}$S$_{1/2}(F=0)$ is about 37pW.

\begin{figure}[!h]
\includegraphics[width=7.5cm]{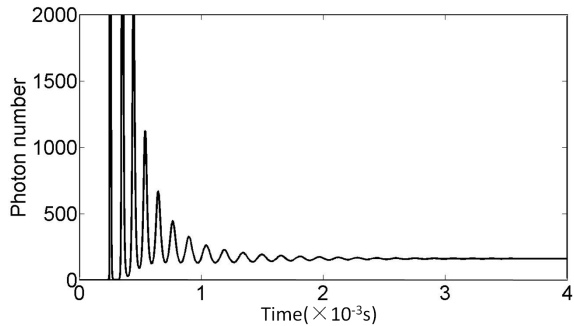}
\caption{The average photon number inside the cavity for
$\Omega_{14}=3.35\times10^{5}s^{-1}$. There are $10^{6}$ ions in the
cavity. The steady-state value of photon number $n$ is 160.}
\label{fig:Fig3}
\end{figure}

The population inversion depends on the Rabi frequency $\Omega_{14}$ and $\Omega_{23}$.
Here we consider the effect of $\Omega_{14}$. The enhancement of $\Omega_{14}$ will increase the population inversion, thus
increasing the probability of lasing transition between the inverted states.
The steady-state value of photon number for $\Omega_{14}=4.75\times10^{5}s^{-1}$
is larger than that of $\Omega_{14}=3.35\times10^{5}s^{-1}$ in the case of the same total ions number(see Fig.~4).
Given $\Omega_{14}=4.75\times10^{5}s^{-1}$, the largest value of $\Omega_{14}$
we can get under general experimental condition, the steady-state value of photon number $n$ is 336.
The power of this $^{171}$Yb$^{+}$ ion laser is about 77 pW.

\begin{figure}[!h]
\includegraphics[width=7.5cm]{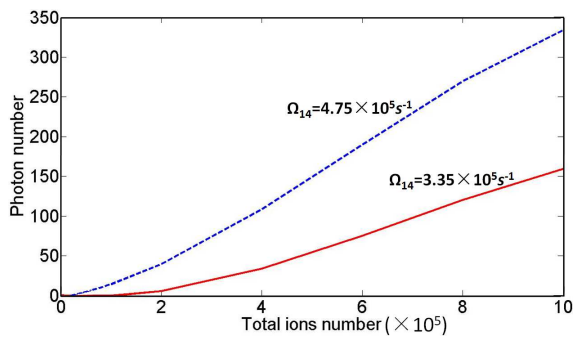}
\caption{(Color online) The steady-state value of photon number $n$
varies with the total ions number $N$ in the cavity. The red line is
for $\Omega_{14}=3.35\times10^{5}s^{-1}$, the blue line is for
$\Omega_{14}=4.75\times10^{5}s^{-1}$.} \label{fig:Fig4}
\end{figure}

As the number of ions in the cavity increases, the steady-state value of photon number is expected to rise accordingly.
The steady-state value of photon number $n$ varying with the total ions number $N$ in the cavity
is also shown in Fig.~4. Considering the actual experimental condition,
we set the largest number of total ions in the cavity to be $10^{6}$.

Compared with the traditional laser, the primary
three conditions are fulfilled for the active optical frequency standards based on
Paul Trap trapped ions just proposed in this paper. The very difference is that
the lifetime of the lasing upper energy level is so long that the natural linewidth
of the laser field is much narrower than the cavity mode linewidth, which is
500 kHz. Besides, the temperature of cold ions trapped in Paul Trap decreases the Doppler
broadening linewidth. Therefore, active ion optical clock operating at the condition of bad cavity
will be a very stable narrow-linewidth laser source since the cavity pulling effect will dramatically
reduce the cavity length noise due to Johnson thermal effect.
It is suitable for single-ion optical frequency standards due to its higher stability.

\section{LINEWIDTH OF ACTIVE YB ION LASER}
Although conventional single-ion optical clocks have reached unprecedented stability and accuracy, the observed linewidth
of the clock transition is limited by the linewidth of the
probe laser~\cite{Chou2,Jiang,Katori,Swallows}. Active ion optical clock, as which we presented in this paper, can offer a new
laser light source for single-ion optical clocks because of its narrow linewidth and high stability.

After laser cooling, the $\gamma=3.1$ Hz natural linewidth clock transition between $^{2}$S$_{1/2}(F=0)$ and
$^{2}$D$_{3/2}(F=1)$ states has been measured with a Fourier-limited linewidth of 30 Hz recently~\cite{Tamm2}. Based on the
mechanism of active optical clock~\cite{Chen1}, to reduce the effect caused by the thermal noise of cavity, the ion laser
cavity should be bad cavity, which means the linewidth of the cavity (or the cavity loss rate) should be much wider than
the linewidth of the gain (the broaden ion transition linewidth after considering the Doppler broadening, collision
broadening and light shift \emph{etc.}). Given the linewidth of ion gain to be 30 Hz as measured, one can set the cavity mode linewidth as
$\kappa=500$ kHz. Then the cavity related noise is reduced by 4 orders of magnitude to below 1 mHz with the best cavity design.
The ion-cavity coupling constant ~\cite{Chen1} can be set around $g=20$ Hz. The quantum-limited linewidth of active ion optical clock~\cite{Yu}
$\gamma_{ionlaser}=g^{2}/\kappa$ is narrower than 1 mHz.

There are still some effects that will shift and broaden the linewidth of active ion optical clock,
like light shift caused by the repumping laser. For $^{171}$Yb$^{+}$ active ion optical clock, the largest light
shift effect is caused by the repumping laser of $^{2}$S$_{1/2}(F=0)$ to $^{2}$P$_{1/2}(F=1)$ transition. By adjusting
this repuming laser at red detuning to $^{2}$P$_{1/2}(F=1)$, its light shift can be greatly reduced by suitable blue detuning to
$^{2}$P$_{1/2}(F=0)$ transition. The repumping laser can be stabilized to an uncertainty of $\Delta=100$ Hz.
Considering that the spontaneous decay rate of $^{2}$P$_{1/2}$ is $2\pi\times23$ MHz, far greater than 100 Hz,
we can write the light shift as $\Delta\nu=2\Omega^{2}_{14}\Delta/\pi\Gamma^{2}_{41}$.
Given the repumping rate $\Omega_{14}=4.75\times10^{5}s^{-1}$, the value
of light shift is 0.69 mHz.

\section{DISCUSSION AND CONCLUSION}
Other ions such as Ba$^{+}$, Ca$^{+}$ and Sr$^{+}$ are also potential candidates for active optical clock and narrow
linewidth ion laser with the same mechanism.

Compared with most clock transitions of traditional passive ion optical frequency standards, which are between
$^{2}$S$_{1/2}$ and $^{2}$D$_{5/2}$ states like Ba$^{+}$, Ca$^{+}$ and Sr$^{+}$,
we recommend the lasing from $^{2}$D$_{3/2}$ to $^{2}$S$_{1/2}$ states for active optical clock and
narrow linewidth ion laser. In this case, here taking the even isotopic $^{88}$Sr$^{+}$ as an example, it can
be a natural three-level laser configuration with the cooling laser used as repumping laser at the same time,
thus simpler than traditional passive ion optical clock. During the 422 nm cooling procedure, ions spontaneously
decay to $^{2}$D$_{3/2}$ state and accumulate at this state with 0.4s lifetime~\cite{Biemont}.
Then, coupled with a bad cavity via stimulated emission, the ions transit from $^{2}$D$_{3/2}$ to $^{2}$S$_{1/2}$ state in the
configuration of three-level ion laser. With a magnetic field along the cavity axis, the laser light will be circularly polarized.

In summary, we propose a scheme of active ion optical clock, i.e.
narrow linewidth bad cavity ion laser with
detailed pumping method, lasing states, output power, linewidth and light shift reduction.
We especially study the $^{171}$Yb$^{+}$ ions in a Paul
Trap and propose to utilize a Fabry-Perot resonator to realize lasing of active optical
frequency standards. The population inversion between
$^{2}$D$_{3/2}(F=1)$ and $^{2}$S$_{1/2}(F=0)$ states can be built up at the time scale of $10^{-3}$s.
The steady-state value of photon number $n$ in the Fabry-Perot cavity increases with $\Omega_{14}$ and the total ions number $N$.
If $\Omega_{14}=4.75\times10^{5}s^{-1}$ and $N=10^{6}$, the steady-state value of photon number is $n=336$
and the laser power is 77 pW. The quantum-limited linewidth of active $^{171}$Yb$^{+}$ ion optical clock
is narrower than 1 mHz. Other ions like Ba$^{+}$,
Ca$^{+}$ and Sr$^{+}$ are suitable for active optical clock. Active optical clock and narrow linewidth
ion laser may also be realized with these candidate ions.

\section{ACKNOWLEDGMENT}
This work is supported by the National Natural Science Foundation of China
(Grant No. 10874009).


\begin{references}

\bibitem{Diddams} S. A. Diddams \emph{et al}., Science {\bf 293}, 825 (2001).

\bibitem{Chou} C. W. Chou, D. B. Hume, J. C. J. Koelemeij, D. J. Wineland, and T. Rosenband, Phys. Rev. Lett. {\bf 104}, 070802 (2010).

\bibitem{Rosenband} T. Rosenband \emph{et al}., Science {\bf 319}, 1808 (2008).

\bibitem{Chou2} C. W. Chou \emph{et al}., Science {\bf 329}, 1630 (2010).

\bibitem{Jiang} Y. Y. Jiang \emph{et al}., Nature Photon {\bf 3}, 158 (2011).

\bibitem{Katori} H. Katori, Nature Photon {\bf 5}, 203 (2011).

\bibitem{Swallows} M. D. Swallows \emph{et al}., Science {\bf 331}, 1043 (2011).

\bibitem{Chen1} J. Chen, Chin. Sci. Bull. {\bf 54}, 348-352 (2009).\\e-print arXiv:physics/0512096.

\bibitem{Chen2} J. Chen, X. Chen, Proc. IEEE 2005 FCS. 608-610 (2005).

\bibitem{Zhuang1} W. Zhuang, D. Yu, and J. Chen, Proc. of 2006 IEEE International Frequency Control Symposium 277-280 (2006).

\bibitem{Zhuang2} W. Zhuang, J. Chen, Proc. of 20th EFTF, 373-375 (2006).

\bibitem{Zhuang3} W. Zhuang \emph{et al}., Proc. of 2007 European Freq. and Time Forum \& IEEE Int'l Frequency Control Symposium 96-99 (2007).

\bibitem{Yu} D. Yu, J. Chen, Phys. Rev. A. {\bf 78}, 013846 (2008).

\bibitem{Chen3} J. Chen, Proceedings of the 7th Symposium, 525-531 (2008).

\bibitem{Meiser1} D. Meiser, Jun Ye, D. R. Carlson, and M. J. Holland, Phys. Rev. Lett. {\bf 102}, 163601 (2009).

\bibitem{Meiser2} D. Meiser, and M. J. Holland, Phys. Rev. A. {\bf 81}, 033847 (2010).

\bibitem{Xie} X. Xie, W. Zhuang, and J. Chen, Chin. Phys. Lett. {\bf 27}, 074202 (2010).

\bibitem{Zhuang4} W. Zhuang, J. Chen, Proc. of 2010 IEEE Int'l Frequency Control Symposium 222-223 (2010).

\bibitem{Zhuang5} W. Zhuang, J. Chen, Chin. Phys. Lett. {\bf 28}, 080601 (2011).

\bibitem{Casdorff} R. Casdorff \emph{et al}., Ann. Phys. {\bf 7}, 41-55 (1991).

\bibitem{Tamm1} Chr. Tamm, D. Engelke, and V. Buehner, Phys. Rev. A. {\bf 61}, 053405 (2000).

\bibitem{Tamm2} Chr. Tamm, S. Weyers, B. Lipphardt, and E. Peik, Phys. Rev. A. {\bf 80}, 043403 (2009).

\bibitem{Biemont} E. Biemont \emph{et al}., Eur. Phys. J. D. {\bf 11}, 355-365 (2000).

\bibitem{An} An K., J Korean Phy Soc. {\bf 42}, 1-13 (2003).





\end{references}
\end{document}